\documentclass[twocolumn, groupedaddress]{revtex4}
\pdfoutput=1 
\usepackage{graphicx}
\usepackage{bm}
\usepackage{hyperref}
\usepackage{amsmath,amssymb}
\usepackage{siunitx}
\usepackage[utf8]{inputenc}
\begin{document}

\title{Three-Dimensional Chiral Magnetization Structures in  FeGe Nanospheres}%
\author{Swapneel Amit Pathak}
\author{Riccardo Hertel}%
 \email{riccardo.hertel@ipcms.unistra.fr}
\affiliation{%
 Universit\'e de Strasbourg, CNRS,\\
 Institut de Physique et Chimie des Mat\'eriaux de Strasbourg, UMR 7504, F-67000 Strasbourg, France
}

\date{\today}

\begin{abstract}
Skyrmions, spin spirals and other chiral magnetization structures developing in materials with intrinsic Dzyaloshinsky-Moriya Interaction display unique properties that have been the subject of intense research in thin-film geometries. Here we study the formation of three-dimensional chiral magnetization structures in FeGe nanospheres by means of micromagnetic finite-element simulations. In spite of the deep sub-micron particle size, we find a surprisingly large number of distinct equilibrium states, namely, helical, meron, skyrmion, chiral-bobber and quasi-saturation state. The distribution of these states is summarized in a phase diagram displaying the ground state as a function of the external field and particle radius. This unusual multiplicity of possible magnetization states in individual nanoparticles could be a useful feature for multi-state memory devices. We also show that the magneto-dipolar interaction is almost negligible in these systems, which suggests that the particles could be arranged at high density without experiencing unwanted coupling.

\end{abstract}

\maketitle

\section{\label{sec:introduction}Introduction}

Three-dimensional (3D) magnetization structures on the nanoscale have recently evolved into a very active field of research \cite{fernandez-pacheco_three-dimensional_2017,fischer_launching_2020, skoric_layer-by-layer_2020}, including {\em e.g.}, magnetic structures in complex nano-architectures \cite{keller_direct-write_2018}, and the tomographic reconstruction of 3D magnetic vector fields in nanocylinders \cite{donnelly_three-dimensional_2017}. The 3D magnetization structure has also been studied in the context of non-centrosymmetric materials, yielding a variety of new structures like skyrmion tubes, chiral bobbers and Bloch point structures in helimagnets~\cite{rybakov_new_2015, rybakov_new_2016}. In these material types, however, the impact of 3D nanoscale confinement and finite-size effects on the magnetization states has not yet been investigated in detail. It is known that  helical states and hexagonal skyrmion lattices can develop in two-dimensional, extended thin films~\cite{muhlbauer_skyrmion_2009,yu_near_2011}, and that the additional degree of freedom that is present in thicker films can give rise to complex magnetization configurations such as skyrmion tubes and chiral-bobbers~\cite{rybakov_three-dimensional_2013,rybakov_new_2015,rybakov_new_2016}. Moreover, patterned thin-film elements can host a variety of complex chiral structures~\cite{beg_ground_2015}, including isolated skyrmions~\cite{zheng_direct_2017}, spin spirals, and  ``horseshoe''-type structures~ \cite{karakas_observation_2018}. Previous studies on finite-size effects in skyrmionic magnetic material have addressed the impact of the film thickness or the lateral size of thin-film elements, but were generally restricted to flat geometries. To study the influence of nanoscale 3D confinement on the magnetization states forming in a helimagnetic material, we perform finite-element micromagnetic simulations on FeGe nanospheres.
In spite of the simplicity of the geometrical shape, we find highly complex magnetic structures in such nanospheres, depending on the particle size and the applied field. This complexity results from the inherently chiral magnetic properties of the non-centrosymmetric material and the constraints imposed by the finite size of the sample.

The general problem addressed in this study, {\em i.e.}, identifying the size dependence of the magnetic ground state, has a long tradition in micromagnetic theory and  simulations~\cite{rave_magnetic_1998,rave_magnetic_2000,usov_nonuniform_2001,hertel_finite_2002}. 
The question of how a magnetic structure is affected by the particle size is often related to the concept of the single-domain limit~ \cite{frenkel_spontaneous_1930,neel_proprietes_1947,kittel_theory_1946,brown_fundamental_1968,
cowburn_single-domain_1999,usov_effective_2002,
yamasaki_direct_2003}, {\em i.e.}, the critical size below which the magnetization in a particle remains homogeneous. This, in turn, is connected to the concept of micromagnetic exchange lengths~ \cite{kronmuller_mikromagnetische_1962,kronmuller_micromagnetism_2003}, which provide material-specific estimates of the characteristic size of fundamental magnetic microstructures, like the width of domain walls or the size of magnetic vortex cores. The exchange lengths result from competing interactions in micromagnetics. In particular they describe a balance between the tendency of the ferromagnetic exchange to maintain a homogeneous magnetic state and other energy terms that favor inhomogeneous structures. 
In the case of non-centrosymmetric magnetic materials with intrinsic chiral properties, the long-range helical period $l_d = 4\pi A/\left|D\right|$ \cite{wilhelm_confinement_2012} plays a role similar to the exchange length in ferromagnets. It represents the length of magnetization spirals forming as a compromise between the ferromagnetic exchange and the antisymmetric exchange due to the Dzyaloshinsky-Moriya interaction (DMI). The constant $D$ denotes the strength of the DMI, {\em i.e.}, the tendency to form helical structures, and $A$ is the ferromagnetic exchange constant. The functional form of $l_d$ is different from that of the magnetostatic exchange length $l_s=\sqrt{2A/\mu_0M_s^2}$ ($M_{\text s}$ is the saturation magnetization), because it describes a periodic, constant modulation instead of the usual kink-type transition with a $\tanh$-type profile. Nevertheless, it can be expected to have similar implications on the size-dependence of magnetic structures, namely that chiral and skyrmionic structures develop in particles with sizes exceeding $l_d$ by a sufficient amount. 

\section{Model system and Numerical Method}
We consider spherically shaped nanoparticles of FeGe with particle radius between \SI{40}{\nano\meter} and \SI{100}{\nano\meter}, thereby extending previous studies on the formation of magnetic structures in this material in the case of planar geometries \cite{zhao_direct_2016,beg_ground_2015}. The spherical shape serves as a simple, fundamental example of a 3D geometry that can host different magnetization states. Moreover, such particles traditionally play an important role in determining the size-dependence of magnetic structures \cite{kittel_physical_1949}. FeGe is a well known B20 type non-centrosymmetric ferromagnet with intrinsic (bulk) Dzyaloshinskii-Moriya Interaction interaction (DMI) \cite{bak_theory_1980,ishikawa_helical_1976,yu_near_2011,
zheng_direct_2017,dzyaloshinsky_thermodynamic_1958,moriya_anisotropic_1960,bogdanov_magnetic_2002}. 
The competition between symmetric ferromagnetic exchange interaction and anti-symmetric DMI gives rise to various complex chiral magnetization configurations. 
We use our custom-developed general-purpose 3D finite-element micromagnetic software package~\cite{hertel_large-scale_2019} to investigate the equilibrium magnetization states forming in the presence of such competing interactions within a confined three-dimensional space.

To model the material properties of FeGe, we use $M_\text{s}=\SI{384}{\kilo\ampere\per\meter}$, $A=\SI{8.78e-12}{\joule\per\meter}$, and $D=\SI{1.58e-3}{\joule\per\meter\squared}$. These material parameters yield a long-range helical period \cite{lebech_magnetic_1989} of $l_{d}=4 \pi A/\left|D\right|\simeq\SI{70}{\nano\meter}$ and a magnetostatic exchange length of $l_{ex}=\sqrt{2A/\mu_{0}M_{s}^2}\approx\SI{5.5}{\nano\meter}$.

The micromagnetic model includes exchange, magnetostatic interaction, DMI and Zeeman energy. We assume that the material of the nanospheres is isotropic, and hence neglect the contribution of magnetocrystalline anisotropy. The total energy  thus reads
\begin{align}\label{eq:energy_density}
E=\int & \Big[ A \sum_{i = x,y,z} (\bm{\nabla} m_i)^{2}-\frac{\mu_{0}}{2}M_{\text s}(\bm{H}_\text{d} \cdot \bm{m}) \nonumber \\ 
&+D\bm{m} \cdot (\nabla \times \bm{m})-\mu_{0}M_{s}(\bm{H}_\text{ext} \cdot \boldsymbol{m})\Big]\,{\rm d}V
\end{align}
where $\bm{m}(\bm{x})$ is the unit magnetization vector,  $H_\text{ext}$ represents the externally applied field, and $\bm{H}_d=-\bm{\nabla} u$ is the magnetostatic field, defined as the gradient field of the magnetostatic scalar potential $u$ \cite{hertel_large-scale_2019}. The effective field $\bm{H}_\text{eff}$ is the variational derivative of the local energy density with respect to the magnetization, $\mu_0\bm{H}_\text{eff}=-M_{\text s}^{-1}\cdot(\delta e/\delta \bm{m})$. This effective field is used in the Landau-Lifshitz-Gilbert (LLG) equation \cite{landau_theory_1935,gilbert_phenomenological_2004} to calculate the magnetization dynamics.

To numerically determine equilibrium states of the magnetization we perform simulations starting from a random initial configuration of the magnetization $\bm{M}(\bm{x}$) and integrate the LLG equation in time until a stable, converged state is found. Several runs are performed with different random initial configurations in order to ascertain that the result represents the ground state, and not a metastable state. Our finite-element software computes the partial effective fields of all energy contributions at each time step and performs the time integration of the LLG equation using an adaptive Dormand-Prince scheme. Since we are only interested in the static ground state, we choose a high damping constant in the LLG equation in order to accelerate the calculation and negelct any dynamic process occuring during the relaxation. The time integration of the LLG equation is thus only used as a means to reach a minimum energy state. The magnetostatic field is calculated with a hybrid FEM-BEM algorithm that uses $\mathcal{H}^2$ hierarchical matrices \cite{hertel_large-scale_2019}, allowing for a particularly fast and memory-efficient computation.
The spatial discretization is done using irregular tetrahedral meshes with cell sizes not exceeding $\SI{2}{\nano\meter}$, which is well below the exchange length of the material. A typical mesh used in our simulations contains  approximately $6\times10^5$ elements for a sphere of radius \SI{70}{\nano\meter}.

\section{\label{sec:results}Magnetic Equilibrium States}

By varying the radius of the nanospheres and the external magnetic field (applied along the $+z$-direction) we obtain, for each combination of radius and external field, a minimum energy equilibrium magnetization state. We first describe in detail the different types of states that we observe.  Afterwards, in section \ref{sec:phase}, we discuss their distribution as a function of the external field and the particle size. Although, generally speaking, the modifications that the lowest-energy magnetic structures undergo by changing the size and the external field are not continuous, it is to some extent possible to interpret the appearance of different magnetization states as a gradual evolution occurring as a result of a changing parameter. To describe such an evolution, we discuss the magnetic ground states found at a fixed sphere radius of $\SI{80}{\nano\meter}$ whilst varying the external magnetic field. The resulting magnetization states are arranged in the order of increasing field.

\subsection{\label{sec:helical state}Helical state}

\begin{figure}[h]
\includegraphics[width=\linewidth]{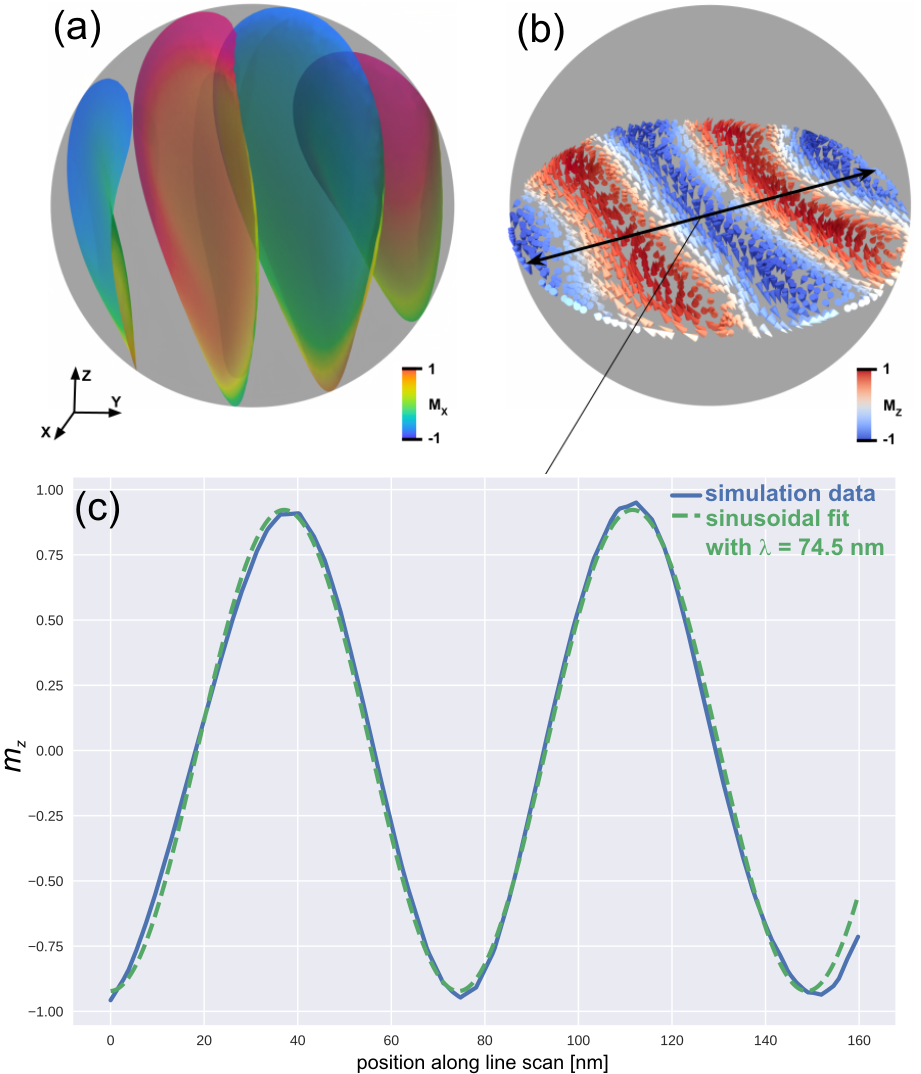}\\
\caption{Helical state in a $r=\SI{80}{\nano\meter}$ FeGe sphere at $H_{\rm ext}=\SI{10}{\milli\tesla}$. The isosurface representation in panel (a) displays the areas where $m_{z}$ is equal to zero.  Panel (b) shows the magnetization configuration on a slice through the equatorial plane, clearly displaying a right-handed magnetization helix. This spin spiral (c) has a wave length of \SI{74.5}{\nano\meter}, which is in good agreement with the analytic long-range helical period of $l_d=\SI{70}{\nano\meter}$ of the material.}
	\label{fig:helical}
\end{figure}

The helical state is characterized by a continuous rotation of the magnetization along an axis perpendicular to the applied field.  The magnetization helix is the direct outcome of the competition between the symmetric ferromagnetic exchange interaction and the anti-symmetric DMI. Alternatively, the arrangement of the magnetization can be interpreted as a periodic sequence of narrow alternating domains, pointing along and opposite to the direction of the external magnetic field, and separated by Bloch walls with the same sense of rotation. These alternating domains can be visualized with the help of iso-surfaces corresponding to $m_{z}=0$, as shown in Fig.~\ref{fig:helical}a.  In this picture, the $m_z=0$ isosurfaces can be regarded as {\em hypothetical} domain walls separating domains aligned parallel and antiparallel to the external field. Since the spatial rotation of the magnetization is rather continuous than localized within domain walls, this interpretation of alternating domains is not strictly correct in micromagnetic terms. Nevertheless, this picture can help to understand the transition towards other states, as described later.

The slice of the magnetization configuration displayed in 
Fig.~\ref{fig:helical}b) shows a right-handed helix extending throughout the sphere, along an axis perpendicular to the external field. One full rotation of this helix occurs on a distance corresponding to the long-range helical period \cite{lebech_magnetic_1989} of the material $l_\text{d}\simeq\SI{70}{\nano\meter}$. The line scan displayed in Fig.~\ref{fig:helical}c) shows that the computed data fits well with the assumption of a perfect spin spiral, with sinusoidal oscillations of the $m_z$ component along the spiral axis. Minor deviations from the ideal value are expected because the analytic calculation of the spin spiral does not consider problem-specific aspects that are included in the simulation, such as the spherical shape, boundary conditions \cite{rohart_skyrmion_2013}, and the magnetostatic interaction. According to our simulations, this helical state is energetically favorable at zero or low external magnetic field, where the exchange energy and DMI dominate.

\subsection{\label{meron state}Meron state}

\begin{figure}[h]
\includegraphics[width=\linewidth]{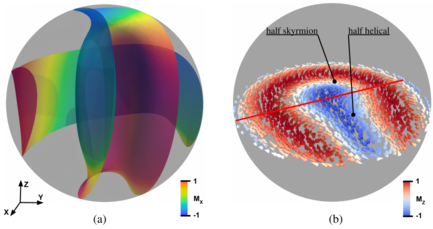}\\
\caption{Three-dimensional meron structure forming as a magnetic ground state in $r=\SI{70}{\nano\meter}$ at $H_\text{ext}=\SI{30}{\milli\tesla}$. The $m_z=0$ isosurfaces are shown in (a). When compared with the isosurfaces of the helical state [cf.~Fig.~\ref{fig:helical}a], one can notice that they display a distinct curvature in the meron state. The magnetization configuration on a central cross-section is shown in panel (b). The state can be interpreted as a combination of a half-helical and half-skyrmion state [cf.~Fig.~\ref{fig:skyrmion}]}.
\label{fig:meron}
\end{figure}

Interpreting the helical state as a magnetic structure with narrow, alternating domains is helpful in order to understand the evolution of the structure as the applied field is increased. Magnetic domain structures react to an increase of the external field such that the domains oriented parallel to the field grow in size, at the expense of domains oriented antiparallel to it. The transition from a helical state to a meron state (Fig.~\ref{fig:meron}) with increasing external field strength can be interpreted in this sense. The increase of favorably oriented regions is recognizable in the isosurface representation, as the previously almost parallel isosurfaces $m_z=0$ bend inwards and connect on one side, cf.~Fig.~\ref{fig:transition1}d,e. 

This field-induced modification of the ground state structure is consistent with a decrease  of the Zeeman energy 
while allowing the magnetic system to preserve to a large extent a spiraling magnetic structure on the length scale $l_d$, as favored by the competition between ferromagnetic exchange and DMI. Note that in ordinary ferromagnets, without DMI, a gradual modification of a periodic domain structure in an increasing external field would occur in a different way, namely by reducing or increasing the distance between neighboring domain walls. Such a domain wall displacement, however, would have a detrimental effect on the periodicity of the spin spirals, and is thus not a viable channel in chiral magnetic materials.

An alternative interpretation of the meron structure consists in considering the magnetization state as a hybrid form of two different chiral structures. More specifically, the magnetization state can be split in two parts (cf.~Fig.~\ref{fig:meron}b), where one half of the nanosphere appears to preserve the structure of a helical state, while the other part displays the charcteristics of a skyrmion,  which will be discussed in the following section. In this sense, the meron state can be considered as an intermediate, transitional structure between these two states.  Meron structures are known from extended two-dimensional system. In such thin films, theory predicts that merons are unstable in isolation, and that instead bi-meron states should form  \cite{ezawa_compact_2011}. However, here, the finite sample size represents a stabilizing factor. We also note that similar examples of isolated meron states have been reported in rectangular shapes \cite{ezawa_compact_2011} and in disc geometries \cite{hrkac_convergent_2017,karakas_observation_2018}, where the structure was denoted as a ``horse-shoe'' state, for obvious reasons.

\subsection{\label{skyrmion state}Skyrmion state}

\begin{figure}[h]
\includegraphics[width=\linewidth]{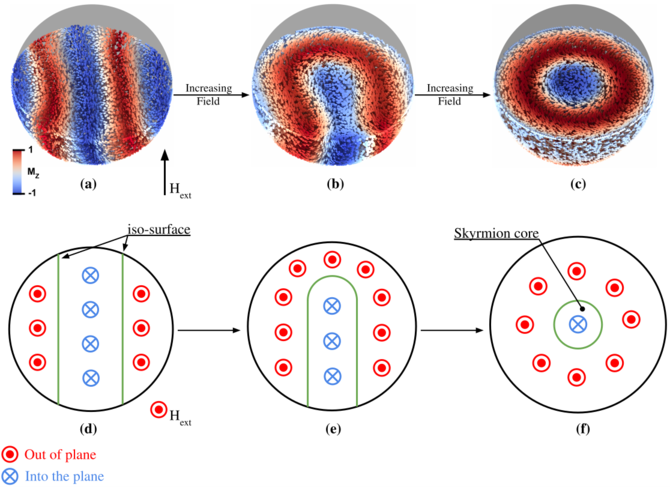}
\caption{\label{fig:transition1} 
Transformation of the magnetic ground state from a helical state (left) into a meron state (middle) towards a skyrmion state (right) as the external field increases. The top row (a), (b), (c) displays simulation results, where the top hemisphere is removed to show the magnetic structure on the central plane. The color code, from blue to red, denotes the magnetization component $m_z$ opposite and along the field direction, respectively. The schematics in the bottom row (d), (e), (f), of the top view, in a simplified way, show the evolution of equilibrium states as the field is increased. The growth of the domains pointing in the direction of the field is not achieved by a reduction of the width of the central domain, but by connecting the iso-surfaces, yielding first the meron state and, at higher fields, the skyrmion state.
}
\end{figure}

Further increasing the external field augments the tendency to expand the regions, or domains, in which the magnetization is aligned along the field direction. This tendency is balanced by the necessity to preserve spin spirals, as required by the interplay of symmetric and antisymmetric exchange. In the isosurface representation, the evolution of a meron state in an increasing external field can be interpreted as a second inwards-bending of the isosurfaces, now connecting the isosurfaces on the opposite side, thereby yielding a circular central core in which the magnetization points opposite to the applied field (Fig.~\ref{fig:transition1}e,f). The resulting axially symmetric configuration is the skyrmion state.

\begin{figure}[h]
\includegraphics[width=\linewidth]{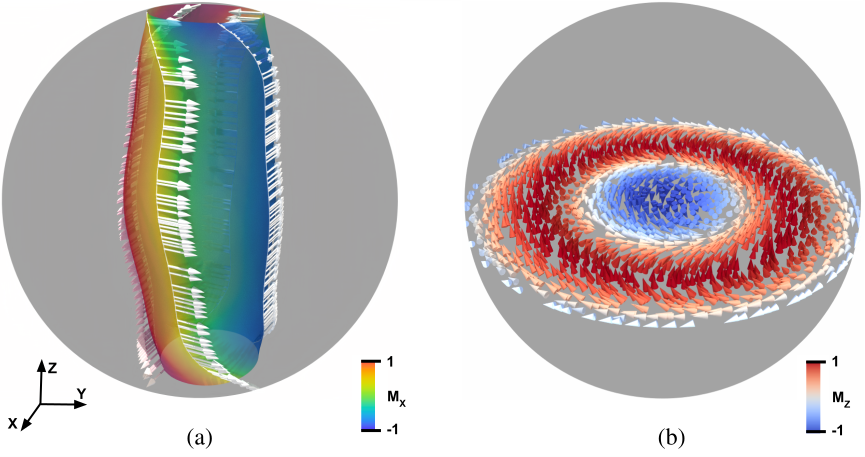}\\
\caption{A three-dimensional skyrmion structure is the magnetic ground state in $r=\SI{80}{\nano\meter}$ at $H_{\rm ext}=\SI{110}{\milli\tesla}$. Panel (a) displays the skyrmion tube in the center, visualized by $m_z=0$ isosurface. It separates the central core from the surrounding circular structure. The skyrmion tube undergoes a twist at the boundaries. This becomes evident by analyzing the change in the position of the magnetic moments pointing in a particular direction, as we move laterally on the skyrmion tube. The magnetic configuration on a horizontal slice in the middle is shown in panel (b), displaying strong similarities with the well-known magnetization texture of a two-dimensional Bloch skyrmion in thin films.}
\label{fig:skyrmion}
\end{figure}

\begin{figure}
\includegraphics[width=\linewidth]{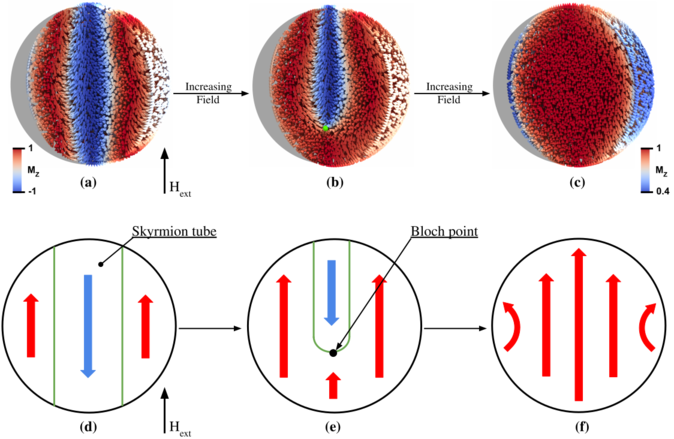}
\caption{\label{fig:transition2}
With increasing field, the skyrmion state (left) transforms first into the chiral-bobber state (middle), which then further evolves into a quasi-saturation state (right). In the simulated structures (a), (b), (c) half of the sphere has been removed to display the evolution and disappearance of the skyrmion tube in the center of the sample. The red and blue color code refers to the value of the magnetization component $m_z$ along and opposite to the field, respectively. The schematics in the bottom row (d), (e), (f) illustrate how the skyrmion core, representing a nano-domain aligned opposite to the field, shrinks as the external field increases. This central domain first becomes smaller as a Bloch point is injected, yielding the chiral bobber state, then it vanishes completely, resulting in a quasi-saturation state with a DMI-induced twist on the surface.
}
\end{figure}

The isosurface representation allows us to clearly visualize the separation of the central skyrmion core from the bulk (Fig. \ref{fig:skyrmion}a). This central, cylindrical region is sometimes referred to as a \textit{skyrmion tube}, or skyrmion line, and it has recently been discussed in the context of high-frequency modes~\cite{lin_kelvin_2019}. The main features of the static configuration are readily recognized by displaying the magnetic configuration on a horizontal slice on the central plane, cf.~Fig.~\ref{fig:skyrmion}b). The magnetization configuration on the central slice shows obvious similarities with the well-known magnetization texture of a two dimensional Bloch skyrmion in a thin film. However, the 3D structure in the sphere has additional features. For instance, the magnetic structure undergoes a twist along the axial direction, as shown in Fig.~ \ref{fig:skyrmion}a), to reduce the DMI energy in the nanosphere. A similar behavior was previously reported by Rybakov et al.~\cite{rybakov_three-dimensional_2013} in the case of thick extended films.

If the external field is further increased, the central core of the skyrmion state pointing in the opposite direction of the field shrinks in size, and the surrounding circular domain oriented along the external field grow. At a certain field, the axially symmetric skyrmion state becomes unstable and transforms into a different magnetization configuration known as a chiral-bobber state~\cite{rybakov_new_2015}. This structure retains to some extent the central skyrmion core, which now however terminates in a Bloch point structure \cite{pylypovskyi_bloch_2012,andreas_multiscale_2014} inside the sphere, cf.~Fig.~\ref{fig:transition2}d,e. The chiral-bobber state can thus be regarded as a hybrid state combining skyrmion and Bloch point structure.

\subsection{\label{chiral-bobber state}Chiral-Bobber state}

\begin{figure}[h]
\includegraphics[width=\linewidth]{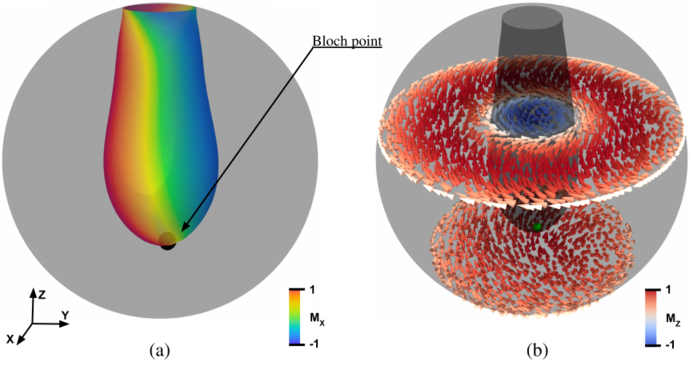}\\
\caption{The chiral-bobber state shown here at  $H_{\rm ext}=\SI{200}{\milli\tesla}$ at $r=\SI{80}{\nano\meter}$, is a complex three-dimensional magnetization structure in which a skyrmion tube terminates in a Bloch point. The conical shape of the residual skyrmion core is visualized by the iso-surfaces corresponding to $m_{z}=0$, shown in panel (a). The magnetization configuration of two slices, one above and one below the Bloch point is shown in (b). On the slice above the Bloch point the structure is similar to the skyrmion state, while below, the magnetization is almost saturated along the field direction.}
\label{fig:chiralbobber}
\end{figure}

To further analyze this magnetic configuration, we display in Fig.~\ref{fig:chiralbobber}b the magnetic structure on two horizontal slices, one above and one below the Bloch point. The configuration on the upper slice resembles that of a skyrmion state, while the one below corresponds to a nearly homogeneous configuration in which the magnetization is largely aligned in the direction of the external field. Chiral-bobber structures have been previously reported, both in theoretical~\cite{rybakov_new_2015} and experimental~\cite{zheng} studies, in thick extended films of non-centrosymmetric ferromagnets. Recently, this magnetization structure has attracted considerable attention as it has been proposed as a candidate for a fundamental unit of information storage, along with the skyrmion state, in future spintronics memory devices \cite{zheng}.

If the external field is further increased, the central core of the chiral-bobber state shrinks in lateral direction until, at a certain field, the Zeeman energy dominates and a quasi-saturated state becomes energetically favorable (Fig.~\ref{fig:transition2}e,f).

\subsection{\label{saturation state}Saturation state}

\begin{figure}[h]
\includegraphics[width=0.7\linewidth]{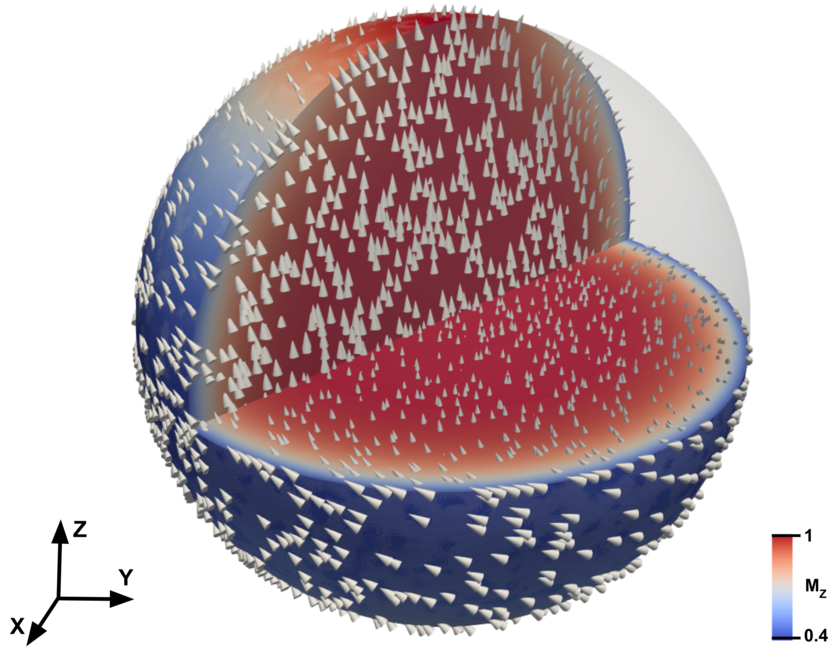}
\caption{Quasi-saturated magnetization state in $r=\SI{80}{\nano\meter}$ at $H_{\rm ext}=\SI{500}{\milli\tesla}$. At such high external magnetic field, the Zeeman energy dominates and the bulk of the magnetization is aligned in the direction of the field. However, a rather significant deviation, which is primarily due to the DMI, occurs at the boundary near the equatorial plane.}
\label{fig:saturation}
\end{figure}

This relatively simple equilibrium state, which is stable at large fields, is characterized by the bulk of the magnetization pointing along the external magnetic field direction, cf.~Fig.~\ref{fig:saturation}. It resembles an ordinary ferromagnetic saturation state. However, near the surface the magnetization deviates, in particular along the equatorial plane. This deviation is primarily due to the DMI, which tends to preserve a chiral structure as far as possible in the presence of a strong external field. The slight curling of the magnetization induced by the DMI is also favored by magnetostatics as the system thereby reduces the magnetostatic surface charges and forms a weakly developed vortex state. Furthemore, the particle surface plays a particular role in the curling of the magnetization due to specific boundary conditions of the DMI interaction~\cite{rohart_skyrmion_2013}, 

\section{\label{sec:phase}Phase diagram}
\begin{figure}[h]
\includegraphics[width=\linewidth]{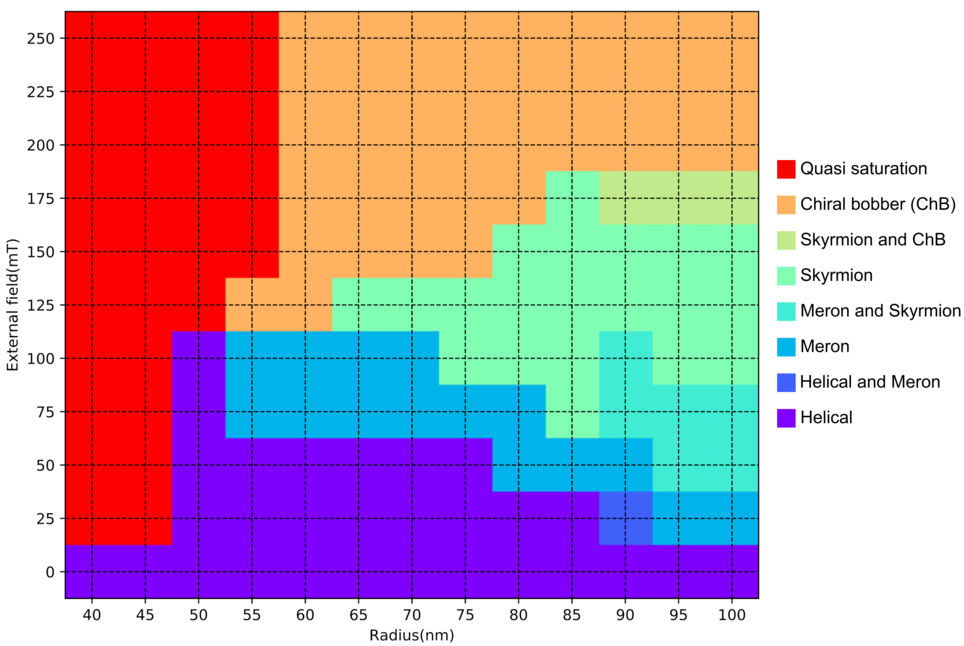}
\caption{Phase diagram of the magnetic ground state of a FeGe nanosphere as a function of the external magnetic field in \si{\milli\tesla} and the radius in \si{\nano\meter}. The different regions outline the parameter ranges in which respective magnetization states represent the lowest-energy configuration. }
\label{fig:phase_diagram}
\end{figure}

In the previous section, we have identified five principal equilibrium states of the chiral magnetization in a FeGe nanosphere, and described their evolution with increasing external field. The stability of these structures, however, also depends on the particle size. To investigate these dependencies, we have performed numerous additional simulations. The numerical results allow us to determine the stability ranges of the five states, as summarized in the phase diagram shown in Fig.~\ref{fig:phase_diagram}. The diagram displays the lowest-energy configuration as a function of the external magnetic field and the radius of the nanospheres.

Remarkably, the skyrmion phase does not exist in FeGe nanospheres below the radius of \SI{65}{\nano\meter}. This size is comparable to the long-range helical period $l_d$ (\SI{70}{\nano\meter}) of the material, which in turn signifies one full rotation of the magnetization. Although there is no direct connection between the structure a spin spiral and the skyrmion state, it is intuitively clear that the sample cannot host a skyrmion structure if it is too small to accommodate two full rotations of the magnetization across the diameter of the sphere.
This trend of disappearing phases continues as we further decrease the radius. Below the radius of \SI{50}{\nano\meter}, the chiral-bobber and meron phase also cease to exist. At this size, the nanosphere diameter approaches $l_d$, and hence, only the helical phase (at lower external fields) and the saturation phase (at higher external fields) are stable. For radii smaller than $\SI{40}{\nano\meter}$, only the saturation phase remains as the particle size falls below $l_d$, leaving no room for even one full rotation of the magnetization.      

A clear distinction of the five principal configurations mentioned above is only possible in particle sizes up to a radius of about \SI{90}{\nano\meter}. In larger nanospheres, hybrid structures appear, which can contain, {\em e.g.}, both a meron and skyrmion structure, or a skyrmion as well as a chiral-bobber. At these larger sizes, the impact of the particle's spherical shape on the magnetic structure diminishes and one observes a gradual transition towards a quasi-continuum of three-dimensional chiral magnetization states, as it would occur in bulk material.

\subsection{\label{demagnetization energy}Impact of magnetostatic interactions}

\begin{figure}[h]
\includegraphics[width=\linewidth]{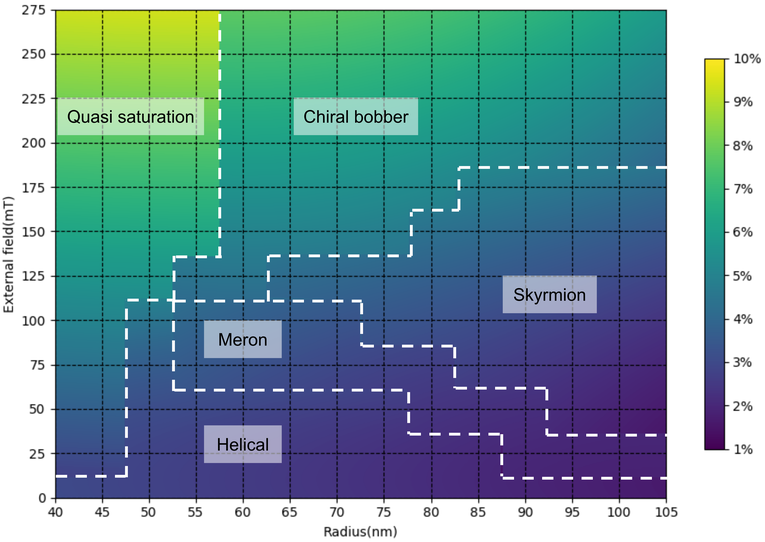}
\caption{\label{fig:dmg_phase}Demagnetization energy as a percentage of the total energy for different magnetic ground states in the phase diagram. In all cases the maximum value remains below 10\si{\percent} through out the phases. The percentage increases only slightly towards the regions of small radius and high fields, where the nanosphere are in the saturation phase.} 
\end{figure}

Having described the various magnetic structure and their formation resulting from the competing interactions of Zeeman energy, ferromagnetic exchange and DMI, we now discuss the impact of the dipolar (magnetostatic) field on these configurations and their distribution. To illustrate the quantitative impact of the dipolar magnetic field, Fig.~\ref{fig:dmg_phase} displays the demagnetization energy as a percentage of the total energy for respective equilibrium states.

It is well known that, in the case of ordinary ferromagnets, the magnetostatic interaction has a decisive impact on the formation of inhomogeneous magnetic structures. The size-dependent equilibrium structure in ordinary ferromagnetic nanoparticles is primarily determined by the balance of the competing interaction of the magnetostatic energy favoring flux-closure states and the ferromagnetic exchange that tends to prevent imohomogeneities of the magnetization. The equilibrium structure is also impacted by the strength of an external magnetic field, and thus the field- and size-dependent distribution of magnetic states in nanoparticles is commonly summarized in phase diagrams similar to ours \cite{rave_magnetic_1998}. However, in our case, the competition is primarily driven, on one side, by the tendency to align the magnetization along the external field direction and, on the other side, by the material's tendency to   develop spiralling magnetization structures on the length scale $l_d$, which in turn is the result of a balance between the ferromagnetic and the antisymmetric exchange interaction. In this latter case, the role of the demagnetizing field is not clear, and it is in fact often neglected in simulations of chiral magnetization structures.

To analyze the impact of magnetostatic interactions on these configurations and distribution, we recalculated the phase diagram by excluding the demagnetization field and energy density from the simulation. Remarkably, we found that this does not alter the results appreciably, yielding in fact essentially the same phase diagram (not shown). This is consistent with the observation that relative impact of the demagnetization energy, displayed in Fig.~\ref{fig:dmg_phase} as the percentage of total energy, is relatively small for all equilibrium states. The demagnetization energy does not exceed \SI{10}{\percent} of the total energy for any of the states. This indicates that, although not strictly negligible, magnetostatic interactions do not play a dominant role in the equilibrium state configuration and distribution. The demagnetization energy becomes only sizable in the upper left part of the plot, {\em i.e.}, towards small radius sizes and high fields, where the particles are in a quasi-saturation state. In  the other equilibrium states, the DMI-induced helical nature of the magnetization structures already reduces the magnetostatic energy by forming states similar to periodically alternating domains, or swirling patterns. The balance between ferromagnetic exchange and DMI thus leads to the formation of magnetic structures which provide a fair amount of magnetic flux closure, so that the demagnetizing energy of the DMI-induced structures remains relatively low. Dipolar fields therefore do not have a decisive impact on helical or chiral magnetization structures. In conclusion, our results indicate that neglecting the magnetostatic interaction is a perfectly acceptable approximation in the simulation of magnetic materials with strong DMI, at least in the case of three-dimensional nanoparticles. This is not necessarily true for flat and thin geometries, where demagnetizing fields generally play a larger role, and where the magnetic surface charges generated by chiral structures have a stronger impact on the total demagnetizing energy. Moreover, the reduced dimensionality of thin films may lower the degree by which chiral or helical magnetization states can achieve a partial magnetic flux-closure.

\section{\label{sec:summary}Conclusion}

Using three-dimensional finite-element micromagnetic simulations, we identified a collection of possible magnetization states in FeGe nanospheres and classified them into five principal categories: helical, meron, skyrmion, chiral-bobber and saturation state. Each of these states can develop as a stable minimum-energy configuration depending on the particle size and external field. This multitude of well-defined magnetic states largely exceeds the variety of magnetic structures that are known from ordinary ferromagnetic nanoparticles, where the size and field dependent variations of the magnetic structure are typically limited to a transition from a homogeneous state to a flux-closure vortex state. In contrast to this, the rich spectrum of magnetic structures in FeGe nanospheres offers the possibility to switch between distinctly different states and thus bears interesting potential for applications as nanoscale multi-state data storage units. 

We note that the geometric confinement, provided by the spherical shape of the nanospheres, allows for the formation of the respective states in \textit{isolation}. This is different from extended films where, for instance, chiral bobber states or skyrmions neither develop as isolated states nor at well-defined positions within the film. The formation of such structures in individual nanoparticles, with well-defined position and orientation, makes it possible to directly address these magnetic structures, thereby opening a pathway towards further investigations on their individual static and dynamic properties.

Our simulations have further allowed us to analyze the influence of magnetostatic interactions on the formation of the magnetic equilibrium configurations and their distribution within a phase diagram. Unlike ferromagnetic particles, where demagnetizing fields decisively impact the equilibrium states, we find that magnetostatic fields only play a negligible role in the case of chiral magnetic structures in FeGe nanospheres. This suggests that, when simulating three-dimensional structures with strong DMI effects, it is acceptable to omit dipolar fields, whose calculation is usually the most expensive part in numerical terms.

\section*{Acknowledgments}
The authors acknowledge funding from the IdEx Unistra through the French National Research Agency (ANR) as part of the Investments for the Future program.


\end{document}